\documentclass{article}
\makeatletter

\@addtoreset{equation}{section}
\makeatother
\usepackage{epsf}
\begin{document}
\baselineskip 12pt
\title{Scalar Dipion States Produced in Heavy Quarkonium  Decays \\
and \\
the Final State Interaction}

\author{M. Uehara\thanks{e-mail: ueharam@cc.saga-u.ac.jp}\\
Takagise-Nishi 2-10-17, Saga 840-0921, Japan}
\date{ }
\maketitle
\begin{abstract}
We study phenomenologically the invariant mass spectra of scalar 
dipion states produced in cascade decays of excited $\Upsilon$ and 
$\psi$, and $J/\psi$ decays into light vector mesons.  The 
dipion production amplitude is given as a product in the Born term 
and an effective scalar form factor written in terms of a unitarized 
chiral theory. The model reproduces the experimental mass 
spectra  very well.
 \end{abstract}
\def\beq{\begin{equation}}　\def\eeq{\end{equation}}
\def\beqa{\begin{eqnarray}}　\def\eeqa{\end{eqnarray}}
\def\beqan{\begin{eqnarray*}}　\def\eeqan{\end{eqnarray*}}
\def\ba{\begin{array}}　\def\ea{\end{array}}
\def\noeq{\nonumber}
\def\mpi{m_\pi} \def\fp{f_\pi}　\def\mK{m_K} \def\fK{f_K}
\def\me{m_\eta} \def\fe{f_\eta}　\def\half{\frac{1}{2}}　
\def\der{\partial} \def\vq{\mbox{\boldmath$q$}}
\def\vg{\mbox{\boldmath$g$}}　\def\vt{\mbox{\boldmath$t$}}
\def\vG{\mbox{\boldmath$G$}}　\def\vK{\mbox{\boldmath$K$}}
\def\vS{\mbox{\boldmath$S$}}   \def\vT{\mbox{\boldmath$T$}}
\def\vrho{\mbox{\boldmath$\rho$}}
\def\del{\delta} \def\vQ{\mbox{\boldmath$Q$}}
\def\vp{\mbox{\boldmath$p$}}
\def\im{{\rm IM}} \def\re{{\rm R}}
\def\CF{{\cal{F}}} \def\CT{{\cal{T}}} \def\CS{{\cal{S}}}
\def\hel{{(\lambda,\lambda)}}
\section{Introduction}
In the last year, dipion production in  cascade 
decays of excited $\Upsilon$  and $\psi$ were studied 
by Ishida et al.,\cite{Komada,MIshida} and $J/\psi$ decays into 
$\phi\pi\pi$ and $\omega\pi\pi$ were studied by Meissner and 
Oller.\cite{MOJpsi}
While the former authors used the preexisting $\sigma$ meson 
expressed by the Breit-Wigner formula with appropriate dipion 
backgrounds,  the latter authors regarded the $\sigma$ state as 
the $\pi\pi$ rescattering effect, which cannot be expressed by the 
Breit-Wigner formula. The existence of the $\sigma$ meson near 
500 MeV has long been an unsettled question though 
it is  listed again as the $f(600)$ state  in the PDG 2002.\cite{PDG02}

It is claimed by Ishida et al. that the $\sigma(500)$ meson is more 
clearly seen in production processes such as  heavy quarkonium decays  
than scattering processes, and that production amplitudes should be 
reanalysed independently from scattering amplitudes by taking account of the 
effect of direct $\sigma$ production.\cite{Komada,MIshida}
Contrarily,  Au, Morgan and Pennington 
state that the production amplitudes should be proportional to 
scattering  amplitudes with adjustable functions real on the physical 
cut.\cite{AMP} Thus, this is another unresolved issue. 

Since the dipion production vertex in the heavy vector meson decay is 
the OZI-forbidden,  the dipion state is produced through 
complicated QCD dynamics, in particular by  gluon dynamics, as 
discussed by many authors in the
1980s.\cite{Voloshin,Novikov,Yan,Belanger}
They succeeded in reproducing  the dipion mass spectra of 
$\Upsilon(2S)\to\Upsilon(1S)\pi\pi$ and $\psi(2S)\to J/\psi\pi\pi$ 
decays under the multipole expansion of gluon fields. The dipion
 mass spectrum in the decay $\Upsilon(3S)\to\Upsilon(2S)\pi\pi$  
could be described  by the model. 
However, they could not reproduce the mass spectrum  of
the decay $\Upsilon(3S)\to\Upsilon(1S)\pi\pi$, which shows a double 
peak structure with  a rapid increase at the threshold. It was considered 
that the $\Upsilon(3S\to 1S)$ decay is supplemented with a sequential 
decay mechanism,\cite{Lipkin,Moxhay} but it turned out that 
the sequential decay amplitude is too small to modify the amplitude so 
as to fit the data.\cite{Zhou} From the phenomenological point of view,
their decay amplitudes are within the Born approximation if the 
strong interactions in the final dipion states are not taken into 
account.  

In this paper we propose another approach to describe 
the dipion mass spectra in the heavy quarkonium decays. Since the 
QCD dynamics are very complicated, we start with a phenomenological 
Lagrangian, calculate the production amplitude in the Born 
approximation, and then incorporate 
the rescattering correction into the Born amplitude in order 
for the total production amplitude  to satisfy the unitarity relation. 
The validity of this form of the production amplitude has been shown in the 
dipion production processes by two-photon collisions 
and the radiative $\phi$ meson decays.\cite{MU06,OO,Marco}
In this approach, the $\sigma$ state need not be introduced as the 
preexisting meson but, rather, appears as a rescattering effect in $\pi\pi$ 
scattering, the amplitudes of which  are given by unitarized chiral 
theories starting with the chiral perturbation theory
\,(ChPT).\cite{IAM,OOBS,OOP,MU04}
The unitarized versions of ChPT are expected to be valid up to 1 GeV or more. 
We use here the two-channel 
scattering amplitudes developed in a previous paper,\cite{MU04}
where the amplitudes are calculated with the Oller-Oset-Pel\`aez 
version\cite{OOP} of the inverse amplitude method.
We emphasize that this rescattering effect produces a clear broad 
bump centered at 500 MeV in the $\pi\pi$ scattering cross section 
without the Breit-Wigner formula. This picture of the 
$\sigma $ state is also seen in other papers.\cite{Meissner,CGL,Gardner}  
The $f_0(980)$ state appears as a typical bound state resonance in 
the $K\bar K$ channel in our model, and therefore has a more stronger 
coupling to the $K\bar K$ channel than the $\pi\pi$ channel.\cite{MU04}

We summarize our results here. 
\begin{enumerate}
\item  The Born terms play a crucial role in reproducing the mass 
spectra in the $\Upsilon$ decays, since the effect of the final state 
interaction  is weakly  dependent on the invariant dipion mass below 
600 MeV. The model reproduces the experimental 
data very well.\cite{CLEO94,CLEO99,Crystal}
\item The essential feature of the $J/\psi\to\phi\pi\pi$ decay 
is due to the lack of the Born term, because the direct production 
of $\pi\pi$ accompanied by $\phi$ is a double 
OZI-forbidden process. Only the $f_0(980)$ resonance is clearly 
seen below 1 GeV, as in the experimental data.\cite{DM2Phi}
\item The $J/\psi\to\omega\pi\pi$ decay involves
the sequential decay $J/\psi\to b_1(1235)\pi\to(\omega\pi)\pi$, in addition 
to  the direct $J/\psi\to\omega\pi\pi$ decay.  Though the 
sequential decay amplitudes are restricted to the Born approximation,
the sum of the two amplitudes reproduces a peak at 450 MeV and an 
up-down structure near the $f_0(980)$ resonance, as in the 
experimental data.\cite{DM2Omega} 
 \end{enumerate}

We explain our model and the kinematics in the next section, and discuss 
the dipion mass spectra in the cascade decays of excited heavy vector 
mesons to the ground states in \S 3,  and those in the $J/\psi$ decay to 
light flavor vector mesons in \S 4. Concluding remarks are 
given in the final section.

\section{Model and kinematics}
We explain our model and summarize the kinematics and notation 
by considering the decay process $V_A\to V_B+(\pi\pi)$, where  
the initial\,(final) vector meson has a mass $M_A(\, M_B)$ and 
four momentum $p_A\,(p_B)$, the final pions have momenta 
$q_1$ and $q_2$, the sum of which is denoted  $Q=q_1+q_2$, and the  
square of the invariant mass of the dipion is written  $s=W^2=Q^2$. 
We discuss the process in the dipion rest frame, called  the 
$Q$-frame hereafter, where the spatial momentum $\vQ\,=\,0$. 
In this frame we have $\vp_A=\vp_B\equiv\vp$, and set the 
positive direction of the $z$-axis in the $Q$-frame equal to $\vp$. 
The energies and momenta of the vector mesons are given as 
\beqa
E_A&=&\frac{M_A^2-M_B^2+s}{2W}, \quad E_B=\frac{M_A^2-M_B^2-s}{2W},\\
p&=&\frac{1}{2W}\left[(M_A+M_B)^2-s)((M_A-M_B)^2-s)\right]^{1/2},\\
&&E_AE_B-p^2=\frac{M_A^2+M_B^2-s}{2}.
\eeqa

The spin of $V_A$ is polarized perpendicular to the beam direction in 
the laboratory frame, where $V_A$ is produced at 
rest.\cite{CLEO94,CLEO99}
In order to go to the $Q$-frame from the laboratory frame, we have to 
boost $V_A$ by $\vp$. The boost transforms the polarization vector 
$\epsilon_A^\mu(0,z)$ (say $V_A$ is polarized in the $z$-direction in 
the laboratory frame) to  
\beq
\epsilon_\mu(\vp,z)=
\sum_{\lambda}\epsilon^{(\lambda)}(p){\hat p}^*_{\lambda}(\Omega^*),
\eeq
where $\Omega^*$ is the angles of $\vp$ in the laboratory frame, 
${\hat p}^*_\lambda=p^*_\lambda/p$ with $p^*_\lambda$ being the 
complex conjugate of the 
$\lambda$-th component of $\vp$, and 
$\epsilon^{(\lambda)}_\mu(p)$ is the polarization 
vector of $V_A$  moving with  momentum $\vp$ and  helicity 
$\lambda$ in the $Q$-frame. 
The dipion production amplitude is then written 
\beqa
F^{(\lambda_B)}_{\rm Lab}(s,\Omega,\Omega^*)&=&
\sum_{\lambda_A}\left\{
\epsilon^{(\lambda_A)}_{A\mu}M^{\mu\nu}(s,\Omega)
{\epsilon^{(\lambda_B)}_{B\nu}}^*{\hat p}^*_{\lambda_A}(\Omega^*) 
\right\}
\noeq \\ 
&=&\sum_{\lambda_A}F^{(\lambda_A,\lambda_B)}(s,\Omega)
{\hat p}^*_{\lambda_A}(\Omega^*),
\eeqa
where we assume that the amplitude $M_{\mu\nu}$ does not depend on 
$\Omega^*$ directly, $F^{(\lambda_A,\lambda_B)}$ is the 
dipion production amplitude with the helicities $(\lambda_A,\lambda_B)$ 
in the $Q$-frame, $\Omega^*$ denotes the angles of $\vp_B$ in the 
laboratory frame and $\Omega$ donotes those of $\vq_1$ in the $Q$-frame. 
The invariant mass spectrum is given as 
\beqa
\frac{d\Gamma_{AB}}{dW}&=&\frac{2}{(4\pi)^5}\frac{p^*_B q_1}{M_A^2}
\int d\Omega\int d\Omega^*\sum_{\lambda_B}
\left|F^{(\lambda_B)}_{\rm Lab}(s,\Omega,\Omega^*)\right|^2
\noeq \\
&=&\frac{2}{(4\pi)^4}\frac{p^*_B q_1}{M_A^2}
\frac{1}{3}\sum_{\lambda_A,\lambda_B}\int d\Omega
\left| F^{(\lambda_A,\lambda_B)}(s,\Omega)\right|^2
\eeqa
where $p^*_B$ is the $V_B$ momentum in the laboratory frame
and $q_1$ the pion momentum in the $Q$-frame:
\beqa
p^*_B&=&\frac{1}{2M_A}\left[((M_A+M_B)^2-s)((M_A-M_B)^2-s)
\right]^{1/2},\\
q_1&=&\frac{1}{2}[s-4m_1^2]^{1/2}.
\eeqa
Thus, in order to study the invariant mass spectra we can construct 
decay amplitudes as if the initial vector meson were unpolarized in 
the $Q$-frame. 
The $\Omega^*$ distributions in the laboratory frame could  be left 
at $O(p_B^2/M_B^2)$ in general.  

Let us define the sign and normalization of the production and 
scattering  amplitudes written in  the two-channel formalism consisting  
of the $\pi\pi$  and  $K\bar K$ channels with the subscript 1 and 2, 
respectively. 
The two-channel $S$-wave isoscalar scattering amplitudes $T_{ij}(s)$  are 
defined as 
\beqa
S_{ij}&=&\delta_{ij}-2i\rho_i^{1/2}(s)T_{ij}(s)\rho_j^{1/2}(s), \\
\rho_i(s)&=&\frac{1}{16\pi}\sigma_i(s)\theta(s-4m_i^2)\qquad 
\sigma_i(s)=\sqrt{1-\frac{4m_i^2}{s}}
\eeqa
with $m_1$ ($m_2$) being the pion\,(kaon) mass. 
The unitarity relation is then written 
\beq
{\rm Im}T_{ij}=-\sum_{k=1,2}T^*_{ik}\rho_kT_{kj}. 
\eeq
The $S$-wave isoscalar dimeson production amplitude, denoted $F_i(s)$, 
must satisfy the unitarity relation
\beq
{\rm Im}F_i^\hel=-\sum_{j=1,2}{F^\hel_j}^*\rho_jT_{ji}. \label{Unitarity}
\eeq

The next task is to search for a suitable model guided by the experimental 
data.  We construct the model as follows:
Since the experimental data show that $D$-wave contamination is very 
small,\cite{CLEO94,CLEO99} we search for a model in which  the helicity 
flip does not occur at the vector meson vertex.  
Thus, the heavy vector meson plays like a 
spectator in the $Q$-frame, since  both the momentum and helicity  
are conserved throughout the decay. 
According to the experimental data, we can see that both in the decays 
$\Upsilon(2S)\to\Upsilon(1S)(\pi\pi)$\cite{CLEO99} 
and $\psi(2S)\to J/\psi(\pi\pi)$\cite{Crystal} 
the invariant $\pi\pi$ mass spectra increases very slowly as in 
the $D$-wave production. Contrastingly the decays  
$\Upsilon(3S)\to \Upsilon(2S)$ and $\Upsilon(3S)\to \Upsilon(1S)$ show 
the threshold behavior consistent with the $S$-wave production.\cite{CLEO94}  
In order to describe such various types of threshold behavior we borrow the form 
of the effective Lagrangian from the argument 
of Meissner-Oller,\cite{MOJpsi} and write it as
\beq
L=g_0V_A^\mu V_{B\mu}S(x)+g_1V_A^\mu V_{B\mu}\der^\nu\der_\nu 
S(x)+g_2V_A^\mu V_B^\nu\der_\mu\der_\nu S(x),
\eeq
where $S(x)$ is the scalar current expressed in terms of the dipion and 
$K\bar K$ fields, and the uncorrelated part of $S(x)$ could be approximated as 
$\sum_i\pi_i(x)^2$ or $K\bar K(x)$. 
We note that all of the pieces of $L$ are discussed in Ref. 3),
and that the resultant production amplitude 
with the $g_0$ term turns out to be similar to theirs. 

The isoscalar Born amplitude is derived using the uncorrelated part of 
$S(x)$ in the Lagrangian as 
\beq
B^{\lambda_A\lambda_B}=(\epsilon_A^{\lambda_A}\cdot
{\epsilon_B^{\lambda_B}}^*)(g_1Q^2-g_0)+
g_2(\epsilon_A^{\lambda_A}\cdot Q)({\epsilon_B^{\lambda_B}}\cdot Q)^*.
\label{Born}
\eeq
Note that the Born amplitude does not depend on $q_i$ but on $Q$, and thus 
the helicity flip is forbidden in the $Q$-frame.  Taking one of the 
constants, say $g_1$, as the overall 
normalization constant, the shape is determined by  the two ratios, 
$g_0/g_1$ and $g_2/g_1$. 

In order to take into account the correlation between the dipion fields  
through the final state interaction and to construct the 
total production amplitude $F_i(s)$, we consider the following  solution of 
the unitarity equation Eq.(\ref{Unitarity}) 
\beq
F^\hel_i(s)=B^\hel_i(s)+\sum_{j=1,2}T_{ij}I^\hel_j(s),
\label{solution}
\eeq
where $B^\hel_i$ is the Born term of the $i$-th channel, but we omit 
the subscript $i$ hereafter, since $B_1^\hel(s)=B_2^\hel(s)$, 
and $I^\hel_j$ is the loop integral connecting the production vertex to 
an $S$-wave point vertex.
The form of the loop integral $I^\hel_i$ in general depends on the 
momentum dependence  of the initial vertex and satisfies the unitarity 
relation 
\beq
{\rm Im}I^\hel_i(s)=-\rho_iB^\hel(s).
\eeq 
Since the Born term depends only on $Q$, $I^\hel_i(s)$ is factorized into 
the Born term and a two-point loop integral as 
\beq
I^\hel_i(s)=B^\hel(s)J_i(s,\mu)
\eeq
with
\beq
J_i(s,\mu)=\frac{1}{(4\pi)^2}\left\{-1+\log\left(\frac{m_i^2}{\mu^2}\right)+
\log\left(\frac{\sigma_i(s)+1}{\sigma_i(s)-1}\right)\right\},
\eeq
where $J_i(s,\mu)$ is the loop integral used in ChPT, and we choose 
the scale parameter $\mu$ to be equal to the value 
$\sqrt{e}\times 0.6=0.989$ GeV used in Ref.19).  
Thus, the total production amplitude is simply written as the factorized 
form, 
\beq
F_i^\hel(s)=B^\hel(s)\left(1+\sum_{j=1,2}J_j(s)T_{ji}(s)\right).
\label{ProdAmp}
\eeq 
Since ${\rm Im}J_i(s)=-\rho_i(s)$, it is easy to see 
that the amplitude satisfies the unitarity relation Eq.~(\ref{Unitarity}). 
It should be noted that the solution Eq.~(\ref{solution}) involves no 
arbitrary parameters other than the three constants involved in the 
Lagrangian.

\section{Cascade decays of heavy vector mesons}
We consider  the four types of cascade decays: 
$\Upsilon(2S)\to\Upsilon(1S)(\pi\pi)$, $\Upsilon(3S)\to\Upsilon(1S)(\pi\pi)$, 
$\Upsilon(3S)\to\Upsilon(2S)(\pi\pi)$, and $\psi(2S)\to J/\psi(\pi\pi)$.  
We denote them as $\Upsilon_{ij}$ and $\psi_{21}$, respectively. 
The clear threshold suppression of the invariant dipion spectra 
is seen in the $\Upsilon_{21}$ and $\psi_{21}$ decays, the threshold  
rising in $\Upsilon_{31}$ and the intermediate behavior in $\Upsilon_{32}$.

The Born terms specified by the helicity are  given as 
\beqa
B^{(0,0)}&=&-\frac{M_A^2+M_B^2-s}{2M_AM_B}(g_1s-g_0)\nonumber\\
&&\hspace{2cm}+
\frac{(M_A+M_B)^2-s}{4M_AM_B}g_2((M_A-M_B)^2-s),\\
B^{(+,+)}&=& -(g_1s-g_0),
\eeqa
where the second term of $B^{(0,0)}$ comes from the last term of 
Eq.(\ref{Born}).
 \begin{figure}
 \epsfxsize=12cm
 \centerline{\epsfbox{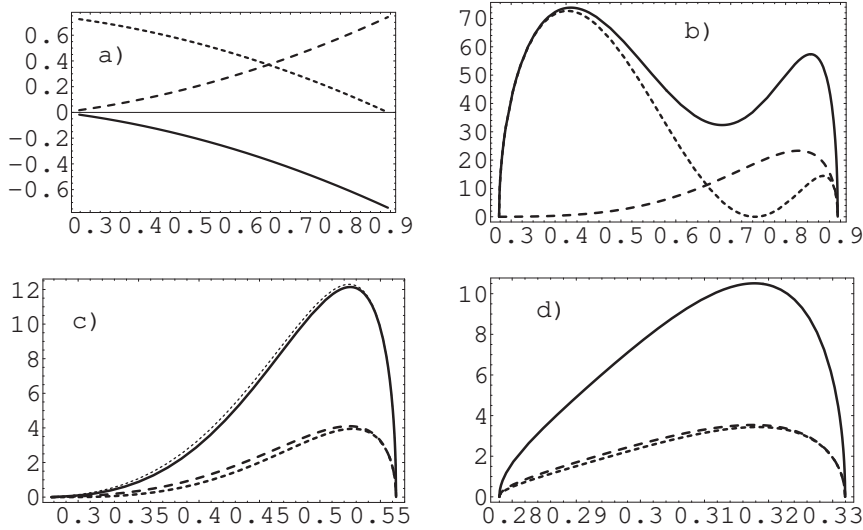}}
 \label{fig:born}
 \caption{a) Born amplitudes with the  ratio $g_0/g_1=3.11m_1^2$ and 
 $g_2=g_1=1$ for the $\Upsilon_{31}$ mass relation. 
 The solid line is the first term of $B^{(0,0)}$, the dotted 
 line the second term, and the dashed line $-B^{(+.+)}$. 
 b) The mass spectrum of $\Upsilon_{31}$ with $g_2/g_1=2$ and 
 $g_0/g_1=3.11m_1^2$. The dotted line is the longitudinal helicity part, 
 the dashed line the transverse helicity part, and the solid line the total cross 
 section. c)  $\Upsilon_{21}$ with the same  $g_0$ and $g_1$ 
 but with $g_2/g_1=0.1$. The dotted line near the solid line is the 
 Voloshin-Zakharov formula with  $\lambda=3.11$. 
 d) $\Upsilon_{32}$ with the same $g_2/g_1$  
but $g_0/g_1=2m_1^2$.  The scales of all of the ordinates are arbitrary, 
and those of the abscissae are GeV.}
\end{figure}
Since the vector meson masses $M_A$ and $M_B$  are much larger than the 
maximum of the dipion mass, the coefficients of $(g_1s-g_0)$ and $g_2$ 
are approximately equal to 1. We show the Born amplitudes in Fig. 1. 
Threshold suppression  is realized under the condition 
$0\,<\,g_0/g_1\,<\,4m_1^2$ and $g_2=0$.
 We note that the Voloshin-Zakharov formula\cite{Voloshin} is written 
$F\propto (s-\lambda m_1^2)$ and the fitting the 
 data gives  $\lambda=3.11-3.42$.\cite{CLEO99} 
 The threshold rising and the 
 double peak structure are obtained by adjusting $g_1$ and $g_2$. For 
 example the choice $g_2/g_1=2$ gives the double peaks  at about 
 $W=0.44$ GeV and 0.85 GeV and the minimum at 0.68 GeV. The dominant 
 contribution to the second peak comes from the transverse helicity 
 amplitude, while the first peak is due to the longitudinal helicity ampitude.

Let us examine the total production amplitudes. 
As explained in the preceding section, the production amplitude is 
factorized into the Born and the rescattering terms;
\beqa
F_1^\hel(s)&=&B^\hel(s)\cdot G_1(s),\label{UdecayAmp}\\
G_1(s)&=&1+J_1(s)T_{11}(s)+J_2(s)T_{21}(s), \label{rescattering}
\eeqa
where the $s\bar s$ production rate is taken to be the same as the 
$n\bar n$ one ($n=u$ and/or $d$), except for the mass difference between 
the pion and kaon, as  in the scattering  processes. 
$G_i(s)$ satisfies the same unitarity relation as the scalar form factor, 
\beq
{\rm Im}G_i(s)=-\sum_{j=1.2}G^*_j(s)\rho_j(s)T_{ji}(s),
\eeq
and the rescattering correction without the kaon-loop,
\beq
\CS(s)=1+J_1(s)T_{11}(s),
\eeq
is very similar to the existing scalar form factor  such as that in 
Ref.  3), as shown in Fig. 2b). 
Thus, $G_i(s)$ could be regarded as an effective scalar form factor 
of the $i$-th dimeson channel. 
The absolute value $\left|G_1(s)\right|$ 
is rather flat for the mass range of the cascade decays as shown in 
Fig. 2a), and therefore the shapes of the mass distributions given by the 
Born terms are not so drastically altered.
\begin{figure}[h!]
 \epsfxsize=12cm
 \centerline{\epsfbox{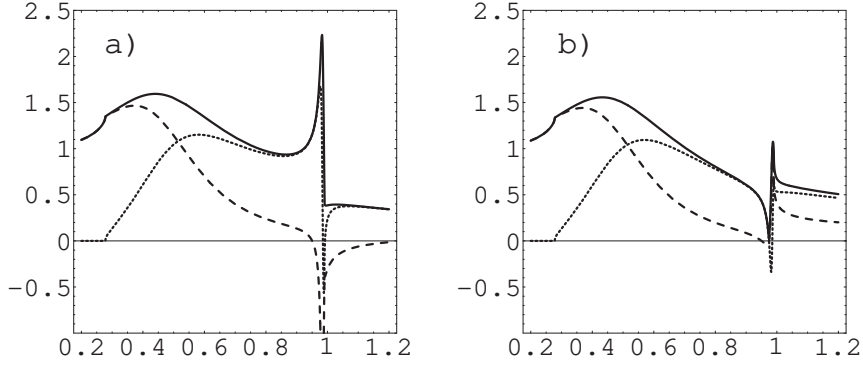}}
 \label{fig:sff}
 \caption{a) The effective form factor $G_1(s)$. The dashed line is the real 
 part, the dotted line the imaginary part, and the solid line the absolute value.
 b) The pionic scalar form factor  $\CS(s)$. The identification  of the lines is 
 the  same as in a).}
\end{figure}

Our model can reproduce the experimental data very well, as shown in 
Figs. 3a) - 3d), though these are not the best fits, where $g_1$ is chosen 
so as to normalize the calculated values to the experimental data. 
The values of $g_n$ are tabulated in Table I.
\begin{figure}[h!]
\epsfxsize=14cm
 \centerline{\epsfbox{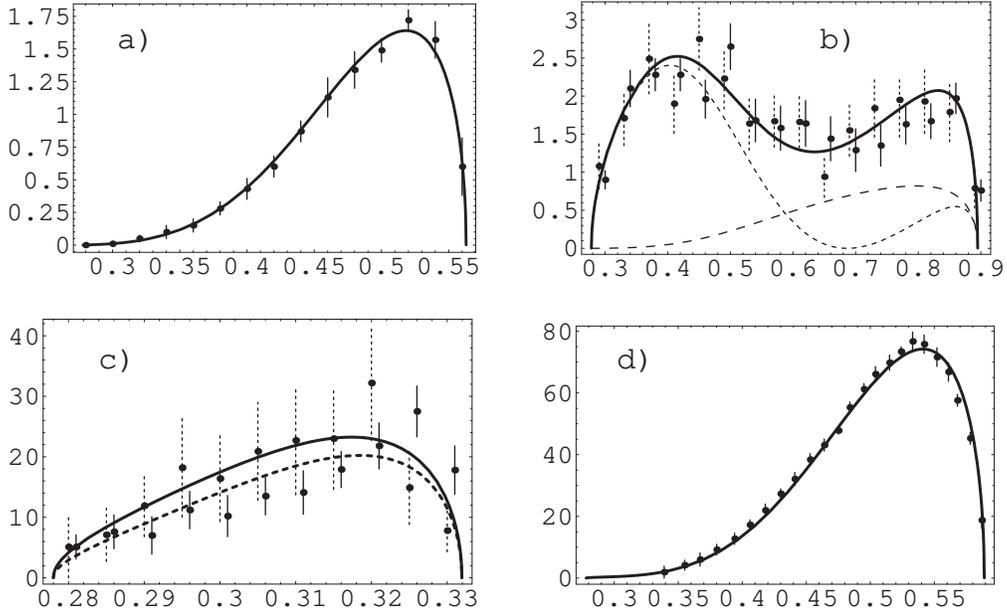}}
 \label{fig:Udecay}
 \caption{The invariant $\pi^+\pi^-$ mass spectra. The experimental 
data in a) - d) were read off of the figures of the published papers 
\cite{Komada,CLEO94,CLEO99} by the present author. The solid and dashed 
error bars in b) and c) correspond to the exclusive and inclusive data, 
respectively. The scales of the ordinates are keV/GeV except for d), where it is 
events/0.01 GeV/c$^2$, and those of the abscissae are GeV.
a) The mass spectrum of $\Upsilon_{21}$ with the data of Fig. 11 in  Ref. 24). 
b) The mass spectra of $\Upsilon_{31}$ with the data of Fig.11 in Ref. 23).
The dotted line is the longitudinal helicity part and the dashed line 
the transverse part. 
c) The mass spectra of $\Upsilon_{32}$ with the data of  Fig.15  in Ref. 23). 
The lower line is for the ratio $g_0/g_1=2m_1^2$, and the upper line for 
the  ratio $0$.  The normalization of the upper and lower lines are different. 
 d) The mass spectrum of $\psi_{21}$ with  the data of Fig. 2(d) in Ref. 1).  }
\end{figure}
\begin{table}[h!]
\caption{$g_n$ for Fig.~3. The left column of $\Upsilon_{32}$ corresponds to 
the lower dotted line and the right one does to the solid line. }
\label{tab:gn}
\begin{center}
\begin{tabular}{|c|r|r|r|r|r|}\hline
$g_n$&$\Upsilon_{21}$&$\Upsilon_{31}$&
\multicolumn{2}{c|}{$\Upsilon_{32}$}&$\psi_{21}$\\ \hline
$g_0/g_1$&$3.11m_\pi^2$&$3.11m_\pi^2$&$2m_\pi^2$&$0m_\pi^2$&
$3.42m_\pi^2$\\ \hline
$g_2/g_1$ &0.1&1.25&0&0&0.2\\ \hline
$g_1$&7.29&3.65&{\small $2.35\times 10^2$}&{\small $1.60\times 10^2$}
& \\ \hline
\end{tabular}
\end{center}
\end{table}

For $\Upsilon_{21}$ and $\psi_{21}$,  $g_0/g_1$ is required to be near to 
the value of Voloshin and Zakharov, and $g_2/g_1$ is required to be small. 
Then, we tentatively set $g_0/g_1=3.11m_1^2$ for $\Upsilon_{21}$ and 
$3.42m_1^2$ for $\psi_{21}$, and $g_2/g_1=0.1$ for $\Upsilon_{21}$ 
and $0.15$ for $\psi_{21}$. The small differences among them should not 
be taken seriously at present. In order to give the rapid rising at the 
threshold in $\Upsilon_{31}$, we need  a large contribution from the
$g_2$ term; we set $g_2/g_1=1.25$ with $g_0/g_1=3.11m_1^2$. The 
helicity structure of the double peaks is the same as in the Born 
approximation. The shape of $\Upsilon_{32}$ is rather sensitive to the 
value $g_0/g_1<2m_1^2$ under  a small value of $g_2$. Since the 
difference between the experimental exclusive and inclusive data is not 
small, we try  two cases that $g_0/g_1=2m_1^2$ and $0$ in Fig. 3c)
It is seen that the values of the two coupling constant ratios do not 
deviate from those of the Born approximation, as stated above.

\section{$J/\psi$ decays to $\phi(\pi\pi)$ and $\omega(\pi\pi)$}
The experimental dipion mass spectrum of the final $\phi\pi\pi$ state 
is much different from that of the $\omega\pi\pi$ state; there is seen 
a clear peak of the $f_0(980)$ resonance in the former,\cite{DM2Phi}
while we see a large broad peak centered at about 450 MeV and a small 
dip-bump structure at the $K\bar K$  threshold in the latter.\cite{DM2Omega}
Since the charm quark 
lines annihilate and an $s\bar s$ or $n\bar n$ pair is created at the 
vector meson vertex in the $J/\psi$ decay into a light flavor vector 
meson, the validity of the spectator assumption is doubtful. 
Nevertheless, we use the same Lagrangian and the same model of the 
final state interaction, but we take account of the sequential decay 
process through the axial vector meson $b_1(1235)$, 
$J/\psi\to b_1\pi\to \omega(\pi\pi)$.

If we consider only the $g_0$ term as in Ref. 3), the resultant 
amplitudes are proportional to $G_1(s)$ for the $J/\psi\to\omega\pi\pi$ 
decay, and to $J_2(s)T_{21}$ for the $\phi\pi\pi$ decay.  
Both amplitudes, however, seem to give too large contributions in the 
$f_0(980)$ resonance region,  so we introduce the $g_1$ term to 
reduce the 
contributions near 1 GeV.  We tentatively use the parameter values, 
$g_1/g_0= 0.7\,{\rm GeV}^{-2}$ and $g_2=0$. 
The overall normalization is controlled by the constant $g_0$.
The Born terms in this parameter set generate a broad peak centered 
near 500 MeV and suppress the spectrum near 1 GeV for the 
$\omega\pi\pi$ decay, which are favorable effects for the model. 
However,  they also produce another large peak near 2 GeV. It turns out, 
however, that the rescattering correction, $G_1(s)$, suppresses 
the second peak drastically, though the  rescattering correction 
cannot be trusted at such higher energies. Thus, the validity of 
the model should be guaranteed for low energy $S$-wave production below 
1.2 GeV or less, but the uniqueness of the parameters cannot be guaranteed. 

\subsection{$J/\psi\to\phi(\pi\pi)$}
The dipion mass spectrum of this decay is similar to that of the 
radiative $\phi$ meson decay,\cite{MU06}, and  the prominent peak is 
seen at about 1 GeV as the $f_0(980)$ resonance.\cite{DM2Phi}
If the $\phi$ meson is the pure $s\bar s$ state, the direct 
$\pi\pi$ production accompanied by $\phi$ is the double OZI breaking 
vertex. The two channel production amplitudes with the lowest order OZI 
breaking vertex are, therefore, given as
\beqa
F_1^\hel(s)&=&B^\hel(s)J_2(s)T_{21}(s),\\
F_2^\hel(s)&=&B^\hel(s)\left(1+J_2(s)T_{22}(s)\right),
\eeqa
where we notice that $F_1$ does not have the Born term.

If the double OZI breaking interaction is allowed, the above formula are  
modified as
\beqa
F_1^\hel(s)&=&B^\hel\left(J_2T_{21}+\alpha(1+J_1T_{11})\right),\\
F_2^\hel(s)&=&B^\hel(s)\left(1+J_2(s)T_{22}(s)+
\alpha J_1(s)T_{12}(s)\right)
\eeqa
where $\alpha$ denotes an additional OZI breaking rate. The calculated 
result on the $\pi^+\pi^-$ invariant mass spectrum is 
given in Fig. 4, where we set 
$\alpha=0.2$, and the overall normalization is fixed at the $f_0(980)$ 
peak. If the double OZI breaking interaction is forbidden, that is $\alpha=0$, 
the small enhancement disappears. 
The too steep rising of the $\pi\pi$ mass spectrum  at the $f_0(980)$ 
resonance is  due to the defect of our $T_{11}(s)$ given in Ref. 19), 
where the phase shift $\delta_{00}$ increases too rapidly near the 
$K\bar K$ threshold. This defect also induces a too sharp and large peak 
in $T_{12}$.
\begin{figure}[h!]
\epsfxsize=7cm
 \centerline{\epsfbox{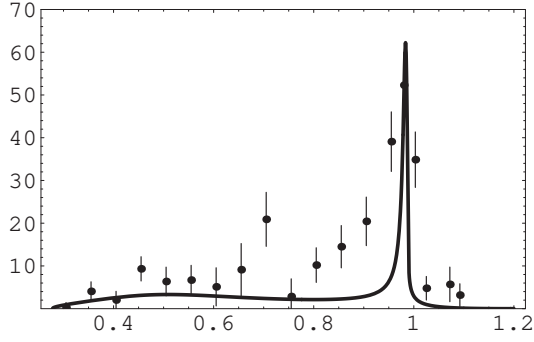}}
 \label{fig:JpsiPhi}
 \caption{The $\pi^+\pi^-$ mass spectrum for $\alpha=0.2$. The 
 experimental data were read off of Fig. 7 in Ref. 3), where the scale of the 
 ordinate is events/25 MeV.}
 \end{figure}

 \subsection{$J/\psi\to\omega(\pi\pi)$}
 While the branching ratio of 
 $J/\psi\to\omega\pi^+\pi^-$ is $(7.2\,\pm\,1.0)\times 10^{-3}$ 
 and that of $\omega\pi^0\pi^0$ $(3.4\pm 0.8)\times 10^{-3}$, 
 the ratio of  $J/\psi\to b_1^\pm(1235)\pi^\mp$ is 
 $(3.0\,\pm\,0.5)\times 10^{-3}$ and that of $b_1^0(1235)\pi^0$ 
 $(2.3\pm 0.6)\times 10^{-3}$, according to the  PDG data.\cite{PDG02}
 This implies that the sequential decay 
 $J/\psi\to b_1\,\pi\to\omega\,\pi\pi$ cannot be ignored. We include both 
 the direct $\omega\,\pi\pi$ decay and the sequential decay 
 process in the analysis. 
 
 The direct production amplitude is given under the same spectator 
 hypothesis as 
 \beq
 F_{\rm dir}^\hel(s)=B^\hel(s)\left(1+J_1(s)T_{11}(s)+J_2(s)
 T_{21}(s)\right),
 \eeq
 where we note that $K\bar K$ production accompanied by $\omega$ does 
 not further break the OZI rule. We show the mass spectra obtained from 
 the direct production amplitude in Figs. 5a) and b) by the dotted lines. They are 
 identical except for the normalization. The spectrum shows a broad peak near 
 450 MeV and the up-down structure near the $K\bar K$ threshold.

 The sequential decay amplitude is calculated as follows:
 We set the effective Lagrangian to describe the vertices 
 $J/\psi\to b_1\pi$ and $b_1\to\omega\pi$ as 
 \beq
 L_{\rm int}=g_{\psi b_1\pi}\psi_\mu b_i^\mu\pi_i(x)+
 g_{b_1\omega\pi}\omega_\mu b^\mu_i\pi_i(x), \label{B1}
 \eeq
 where $\psi_\mu(x)$($b_i^\mu(x)$, $\omega_\mu(x)$) is the 
 $J/\psi$($b_1$, $\omega$) field, and 
 we assume that both of the decays arise through a pure $S$-wave at the  rest 
 frame of the  mother particle. We write the Born term for the sequential 
 decay as 
 \beq
 B_{\rm seq}=g_{\psi b_1\pi} g_{b_1\omega\pi}
 \left\{
 \frac{(\epsilon_\omega^*\epsilon_b)(\epsilon_b^*\epsilon_\psi)}
 {P_b^2-M_b^2+iM_b\Gamma_b} +
 \frac{(\epsilon_\omega^*\epsilon_b)(\epsilon_b^*\epsilon_\psi)}
 {{P'_b}^2-M_b^2+iM_b\Gamma_b}\right\},
 \eeq
 where $M_b$ and $\Gamma_b$ are the mass and total width of $b_1$, and 
 \beqa
 P_b&=&p_\psi-q_2=p_\omega+q_1,\\
 P'_b&=&p_\psi-q_1=p_\omega+q_2.
 \eeqa
 Summing the helicities of the intermediate $b_1$ meson, we have
 \beqa
 B^\hel_{\rm seq}(s,z)&=&g_0R\left\{
 \frac{N^\hel(s,z)}{D(s,z)}+\frac{N^\hel(s,-z)}{D(s,-z)}\right\},\\
 D(s,z)&=&\frac{M_\psi^2+M_\omega^2-s+2m_1^2}{2}-M_b^2+
 iM_b\Gamma_b-2pqz,\\
 N^\hel(s,z)&=&-\left[({\epsilon_\omega^{(\lambda)}}^*\cdot
 \epsilon_\psi^{(\lambda)})
 +(\epsilon_\omega^{(\lambda)}\cdot q_1)^*
 (\epsilon_\psi^{(\lambda)}\cdot q_2)/M_b^2\right]
 \eeqa
with $R=g_{\psi b_1\pi} g_{b_1\omega\pi}/g_0$.
 The sequential decay amplitude includes the $D$-wave contribution, but 
 we calculate only the $S$-wave part in the Born approximation in this 
 paper. 
 \begin{figure}[h!]
\epsfxsize=14cm
 \centerline{\epsfbox{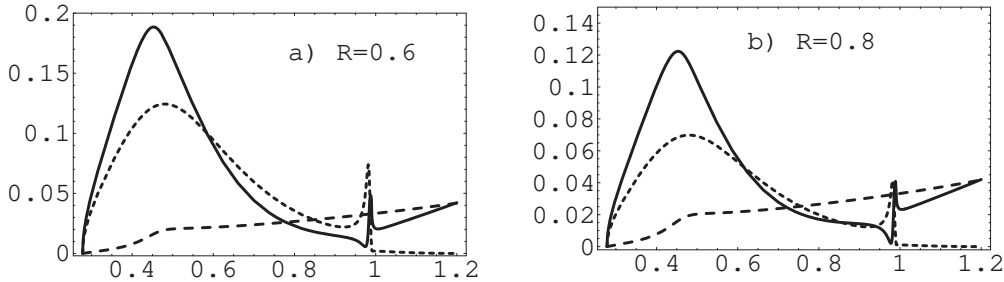}}
 \label{fig:JpsiOmega}
 \caption{The $\pi^+\pi^-$ mass spectrum. The left side is the results for the 
 relative ratio $R=0.6$, and the right for $R=0.8$.
 The spectra calculated with the total amplitude are given by the solid lines, 
 those with the pure direct production amplitude by the dotted lines, and those 
 with  the sequential decay term by the dashed lines. 
  The scale of the ordinate is keV/GeV, using the value of $g_0$ 
 estimated as  explained in the text. }
 \end{figure}
 
 The mass spectrum is calculated with  the total amplitude, 
 \beq
 F_{\rm tot}^\hel=F_{\rm dir}^\hel+B_{\rm seq}^\hel|_S,
 \eeq
 where the second term is the $S$-wave part of the sequential Born term. 
 It should be noted that the tolal amplitude does not satisfy exact 
 unitarity, since the sequential decay amplitude is within the Born 
 approximation. 
 If we compare the decay widths $J/\psi\to (b_1\pi)_{I=0}$ and 
 $b_1\to\omega\pi$  calculated in the Born approximation with the 
 experimental data, we can estimate the product of the two coupling 
 constants as 
 $g_{\psi b_1\pi} g_{b_1\omega\pi}=2.53\times 10^{-2}\,{\rm GeV}^2$,  
 where we take 
 ${\rm Br}(J/\psi\to (b_1\pi)_{I=0})=5.3\times 10^{-3}$ and 
 the $S$-wave partial 
 width of $b_1$ as  $\Gamma_S(b_1\to \omega\pi)=100$ MeV.\cite{PDG02}
 Setting $R=0.6$, then, gives $g_0=4.22\times 10^{-2}\,{\rm GeV}^2$, 
 and $R=0.8$ gives $1.78\times 10^{-2}\,{\rm GeV}^2$. 
 We show the mass spectra calculated with the model using the estimated 
 coupling constants  in Fig. 5, but we do not compare the calculated mass 
 spectrum with  the experimental data,  
 because the data include the $D$-wave contribution which could not be 
 disregarded above 500 MeV. From Fig. 5 we observe that the 
 $S$-wave mass distribution shows a large peak at about 450 MeV,  
 small structure at the $f_0(980)$ resonance and increasing 
 behavior of the sequential decay, all of which are consistent with the 
 experimental data.\cite{DM2Omega} We also see that the interference 
 between the two amplitudes cannot be ignored, but the inclusion of the final 
 state interaction in the sequential decay may alter the pattern of the 
 interference.

 \section{Concluding remarks}
 We have demonstrated the validity of our phenomenological  model 
 in describing the mass spectra of the dipion states produced in heavy 
 flavor vector meson decays. We have shown that the model describes 
 the mass spectra very well with two ratios of the three 
 coupling constants, and an overall normalization. The model takes 
 account of the final state interaction between the produced dipion 
 state through  the two-point loop integral familiar to ChPT and the 
 two-channel $S$-wave scattering amplitudes given by the unitarized 
 chiral theory. 
 The form of the total production amplitudes in our model is not 
 proportional to the scattering amplitudes, as in the so-called universal 
 hypothesis, nor independent of the scattering  processes. The model does not 
 need the Breit-Wigner formula for the bump at 500 MeV and the $f_0(980)$ 
 resonance, neither of which is a  
preexisting meson state but, rather, dynamical objects emerging from  the 
multichannel scattering dynamics. The fact that the $f_0(980)$ peak is 
more clearly seen in production processes than scattering 
processes  is due to the large coupling of the $f_0(980)$ state with the 
$K\bar K$ channel. Such a special effect is not expected for  
the $\sigma$ state, because the $K\bar K\to \pi\pi$ amplitude in this 
region does not have any enhancement. We conclude that these 
decay processes cannot be more useful in revealing the nature of the $\sigma$ 
state than scattering processes deduced from the peripheral 
production processes. We again emphasize that the $I=J=0$ $\pi\pi$ 
scattering amplitude obtained in the unitarized 
chiral theory generates the broad peak near 500 MeV without a 
preexisting $\sigma$ field expressed by the Breit-Wigner formula.\cite{MU04}

Although we reproduced the experimental data well, 
it is impossible to predict the variety of the ratios of the coupling constants 
and the overall normalization constants within our phenomenological model. 
The process dependence of the parameters could be analyzed through 
complicated QCD calculations, but this is far beyond our scope.

The angular distribution of $\vq_1$ in the $Q$-frame indicates that the 
dipion state is not a pure $S$-wave state for $\Upsilon_{21}$ and 
$\Upsilon_{31}$.\cite{CLEO94,CLEO99}
This is also the case for $J/\psi\to\omega\pi\pi$,\cite{DM2Omega}
where the angular distribution for $W\,>\,600$ MeV shows a significant 
angle dependence like a $D$-wave contamination. As pointed out in \S 4, 
the $D$-wave component of the sequential decay 
$J/\psi\to b_1(1235)\pi\to\omega\pi\pi$ would contribute substantially 
to the large tail of the $f_2(1270)$  resonance.

We do not consider the calculation of 
the $K\bar K$ spectra in this paper,  because there are ambiguous  
contamination of multiple sequential decays 
such as $J/\psi\to K^*(1270)\bar K+{\rm c.c.}$ and 
$K^*(1400)\bar K+{\rm c.c.}$ for 
both of the final states, $\phi K\bar K$ and $\omega K\bar K$. 
According to the PDG data\cite{PDG02} the branching ratios are 
$<\,3.0\times 10^{-3}$ for the former and $(3.8\pm 1.4)\times 10^{-3}$ for 
the latter decay channel. Neither of the axial vector mesons can decay 
into the $\phi K$ state, but they can couple to it. The pure direct decay 
component gives  rapid threshold rising  and then decreasing behavior 
in the $K\bar K$ mass distribution.  This is due to the peak behavior of 
$T_{22}$ near the $K \bar K$ threshold owing to the $f_0(980)$ state. 
Similar threshold rising behavior
is seen in $\gamma\gamma\to K\bar K$ and 
$\phi\to\gamma K^0\bar K^0$.\cite{MU06}\\

\section*{Acknowledgements}
The author would like to dedicate this paper for the memory of Professor 
Sakae Saito.  He thanks  the 
Department of Physics, Saga University for the hospitality 
extended to him.

\end{document}